\documentclass[a4paper]{jpconf}
\usepackage{graphicx}
\usepackage{amssymb}
\usepackage{amsfonts}
\usepackage{caption}
\usepackage{multirow}

\usepackage{iopams}

\newcommand{\LMO}{Li$_2$MoO$_4$}
\newcommand{\enLMO}{Li$_2$$^{100}$MoO$_4$}

\newcommand{\enCMO}{$^{48\textrm{depl}}$Ca$^{100}$MoO$_4$}
\newcommand{\zerodbd}{$0\nu\beta\beta$}
\newcommand{\twodbd}{$2\nu\beta\beta$}
\newcommand{\mbb}{$ m_{\beta\beta}$}
\newcommand{\ml}{$m_\mathrm{lightest}$}
\newcommand{\Tzerov}{$T_{1/2}^{0\nu}$}
\newcommand{\gA}{$g_\mathrm{A}$}
\newcommand{\gAeff}{$g_\mathrm{A}^\mathrm{eff}$}

\newcommand{\Mo}[1]{$^{#1}$Mo}

\newcommand{\Ge}[1]{$^{#1}$Ge}
\newcommand{\Xe}[1]{$^{#1}$Xe}
\newcommand{\Se}[1]{$^{#1}$Se}
\newcommand{\Te}[1]{$^{#1}$Te}
\newcommand{\Ca}[1]{$^{#1}$Ca}

\begin{document}
\title{Neutrinoless double beta decay experiment}

\author{Yong-Hamb Kim}

\address{Center for Underground Physics, Institute for Basic Science (IBS), Daejeon 34126,  Korea\\
University of Science and Technology (UST), Daejeon 34113, Korea\\
 Korea Research Institute of Standards and Science (KRISS), Daejeon 34113, Korea
}

\ead{yhk@ibs.re.kr}

\begin{abstract}
The search for neutrinoless double beta decay (\zerodbd) is one of the key experiments for determining unresolved properties of neutrinos. Experimental observation of \zerodbd{} would provide a clear demonstration of the Majorana nature of neutrinos and a lepton number violating process in particle physics. 
This report is a brief review describing the \zerodbd{} process and its significance.
The detector technologies used in the present and proposed \zerodbd{} experiments are compared among their recent experimental results. Moreover,  the sensitivities for \zerodbd{} discovery in a number of present and proposed \zerodbd{} experiments are introduced. 
\end{abstract}

\section{Introduction}

Neutrinoless double beta decay (\zerodbd) is a hypothetical decay process related to the fundamental understanding of particle physics and the universe. This rare decay process can occur only if neutrinos are massive Majorana particles (i.e., their own anti-particles).  Moreover, the decay is allowed only when lepton number conservation is violated in the physical process. In principle, a new understanding of particle physics would need to be introduced to explain lepton number violation (LNV) process beyond the standard model (SM). 

Recent experimental and theoretical studies have revealed various properties of neutrinos that are a class of elementary particles. 
Recent observations of neutrino oscillations~\cite{fukuda1998prl,cleveland1998aj,ahmed2001prl,fogli2012prd} 
have demonstrated that the flavor states are expressed as mixed combinations of mass eigenstates established via a characteristic matrix known as the Pontecorvo-Maki-Nakagawa-Sakata (PMNS) matrix~\cite{maki1962}. 
These results suggest that neutrinos have a nonzero mass. Moreover, the mass eigenstates evolve in time, resulting in oscillating appearance and disappearance of the flavor states.

Experimentally, the square differences of the mass eigenstates have been found from oscillation measurements of atmospheric and solar neutrinos~\cite{fukuda1998prl, cleveland1998aj,ahmed2001prl}. 
 However, their absolute mass scales are still unknown. Moreover, it has not yet been concluded whether neutrinos are Majorana or Dirac particles.  That is a fundamental question linked to an in-depth understanding of particle physics~\cite{avignone2008rmp}.

\section{General aspects of neutrinoless double beta decay}

A normal double beta decay (\twodbd) is a second order weak process in nuclear physics. The two neutrino mode of double beta decay was first predicted from the study of a nuclear structures~\cite{goeppert1935}. The \twodbd{} process  is always associated with the emission of two electrons and two antineutrinos 
expressed as
\begin{equation}
(Z,A)\rightarrow(Z+2,A)+2 \beta^- + 2 \bar \nu_e.
\end{equation}  
About 10 nuclei have been found to undergo double beta decay with a variety of half-lives on the order of $T^{2\nu}_{1/2} \approx 10^{19-24}$~\cite{barabash2015} among the 35 nuclei for which the process is energetically allowed~\cite{tretyak2002}. 

On the other hand, in the case of \zerodbd{} 
no neutrinos are emitted in association with  the decay process, expressed as 
\begin{equation}
(Z,A)\rightarrow(Z+2,A)+2 \beta^- .
\label{equation_0nbb}  
\end{equation}
This hypothetical process violates lepton number conservation by 2. 
The possibility of neutrino-less decay was first discussed independently by E. Majorana and G. Racah~\cite{majorana1937,racah1937}. W.H. Furry made a specific discussion of the neutrinoless process in $\beta\beta$ decay based on the Majorana theory~\cite{furry1939}.

\begin{figure}[t]
\includegraphics[width=15pc]{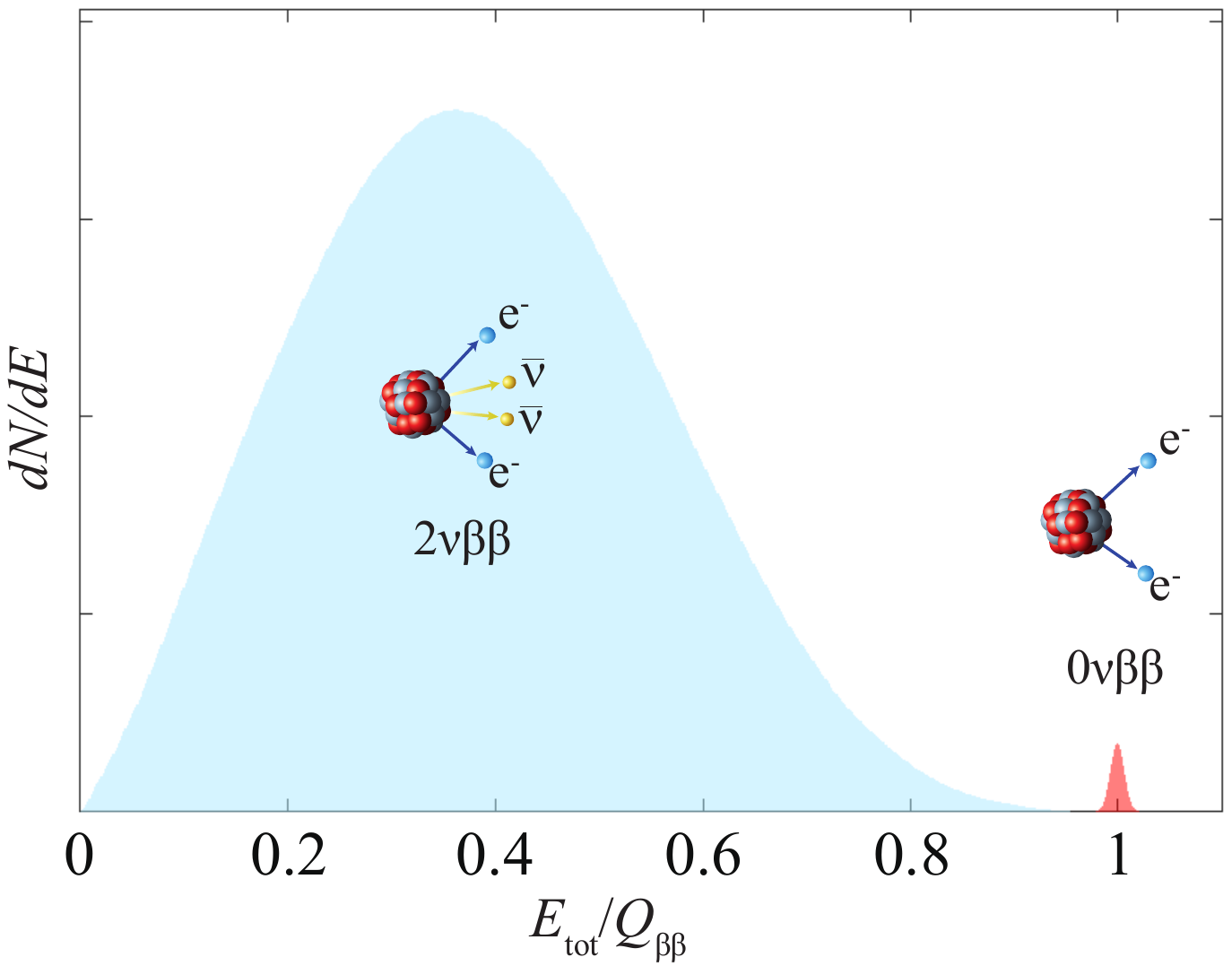}\hspace{2pc}%
\begin{minipage}[b]{16pc}\caption{\label{label} Energy spectra the electrons from the \twodbd{} and \zerodbd{} processes.}
\label{dbd_spec}
\end{minipage}
\end{figure}

Because a \zerodbd{} process is associated with the emission of only two electrons,
the decay energy is shared between the two electrons with a negligible amount by the daughter nucleus.
Consequently, \zerodbd{} events appear as the peak at the end point of a broad electron energy spectrum of \twodbd{} decay as shown in Fig.~\ref{dbd_spec}. 
The \zerodbd{} decay rate is highly exaggerated in this figure. Backgrounds and statistical fluctuations are not included in the spectra.

In general, the decay rate $ \Gamma_{0\nu}$ of the \zerodbd{} process is expressed as 
\begin{equation}
%    \Gamma_{0\nu} = [T^{0\nu}_{1/2}]^{-1} = G_{0\nu} \left| M_{0\nu} \cdot \eta \right|^2
    \Gamma^{0\nu} = 1 / T^{0\nu}_{1/2}  = \sum_{x} G_{x}^{0\nu} \left| M_{x}^{0\nu}  \eta_x \right|^2
\label{equation_eta0nbb}
\end{equation}
where $T^{0\nu}_{1/2}$ is the \zerodbd{} half-life of the element, $x$ and $\eta_x$ denote physics mechanisms and their corresponding functions for LNV processes, and $G_x^{0\nu}$ and $M_x^{0\nu}$  are the phase space factors and the nuclear matrix elements (NMEs) of each mechanism for the decay, respectively~\cite{rodejohann2012jpg}. In the case of the light neutrino exchange model (i.e., a commonly accepted mechanism for interpreting \zerodbd), the decay rate becomes  
\begin{equation}
      1/ T^{0\nu}_{1/2}   = G^{0\nu} \left| M^{0\nu} \right|^2  m_{\beta\beta}^2.
%       \Gamma^{0\nu}    = G^{0\nu} \left| M^{0\nu} \right|^2  m_{\beta\beta}^2.
%    \Gamma^{0\nu}    = G^{0\nu} \left| M^{0\nu} \right|^2 \left< m_{\beta\beta}\right>^2.
\label{equation_decayrate}    
\end{equation}
where \mbb{} is the effective Majorana mass. In this interpretation, no other mechanism contributes to the zero-neutrino mode of $\beta\beta$ decays. The phase space factor $G^{0\nu}$ is exactly calculable in atomic physics~\cite{kotila2012prc}. However, uncertainties by factors of  2--3 are found in the NME $M^{0\nu}$ calculation for all  isotopes of \zerodbd{} candidates~\cite{engel2017rpp}. 
 
 Because the decay rate calculations for the $\beta$ and  \twodbd{} processes are adjusted with the effective axial-vector coupling constant \gA, similar quenching process may exist in \zerodbd. 
It is noted that even with small \gAeff values, \zerodbd{} rates are reduced  by factors of \textit{only} 2--6 because \gA{} is largely incorporated with the NMEs~\cite{suhonen2017prc}. On the other hand, some theoretical predictions suggest no or reduced quenching might be needed in \zerodbd{} because the quenching process depends on the energy scale \cite{dolinski2019}. This subject will require further theoretical and experimental investigations.

 Although \zerodbd{} discovery would be a clear demonstration of Majorana neutrinos and LNV process allowed in particle physics, it  should be noted that Eq.~(\ref{equation_decayrate}) is an interpretation of general LNV mechanisms. Other theoretical models also exist that lead to \zerodbd{} rates expected to be of the same order as the light neutrino exchange mechanism~\cite{deppisch2012jpg}. 
Therefore, full understanding of \zerodbd{} process requires  \zerodbd{} measurements in more than one isotope to clarify the model responsible for \zerodbd.  
Moreover, the uncertainties of the NME calculations  enhance the need for \zerodbd{} search experiments using various nuclei. 

%The NME $M^{0\nu}$ is calculable in nuclear physics with 2--3 times uncertainty for each isotope of \zerodbd{} candidates~\cite{engel2017rpp}.

\section{Neutrinoless double beta decay and neutrino mass}
Because of the virtual neutrino exchange in the standard process, the effective Majorana mass in Eq.~(\ref{equation_decayrate}) is a coherent sum of the neutrino masses expressed as 
\begin{equation}
 m_{\beta\beta}  = \left| \sum U^2_{ei} m_i \right|
\label{equation_mbb}
\end{equation}
where   the  $U_{ei}$ are the elements of the PMNS matrix containing the mixing angles and CP phases, and $m_i$ are the mass eigenvalues.
On the other hand \mbb{} is a measurable quantity of neutrino mass from a half-life measurement  based on the light neutrino exchange model.

\begin{figure}[t]
\includegraphics[width=17pc]{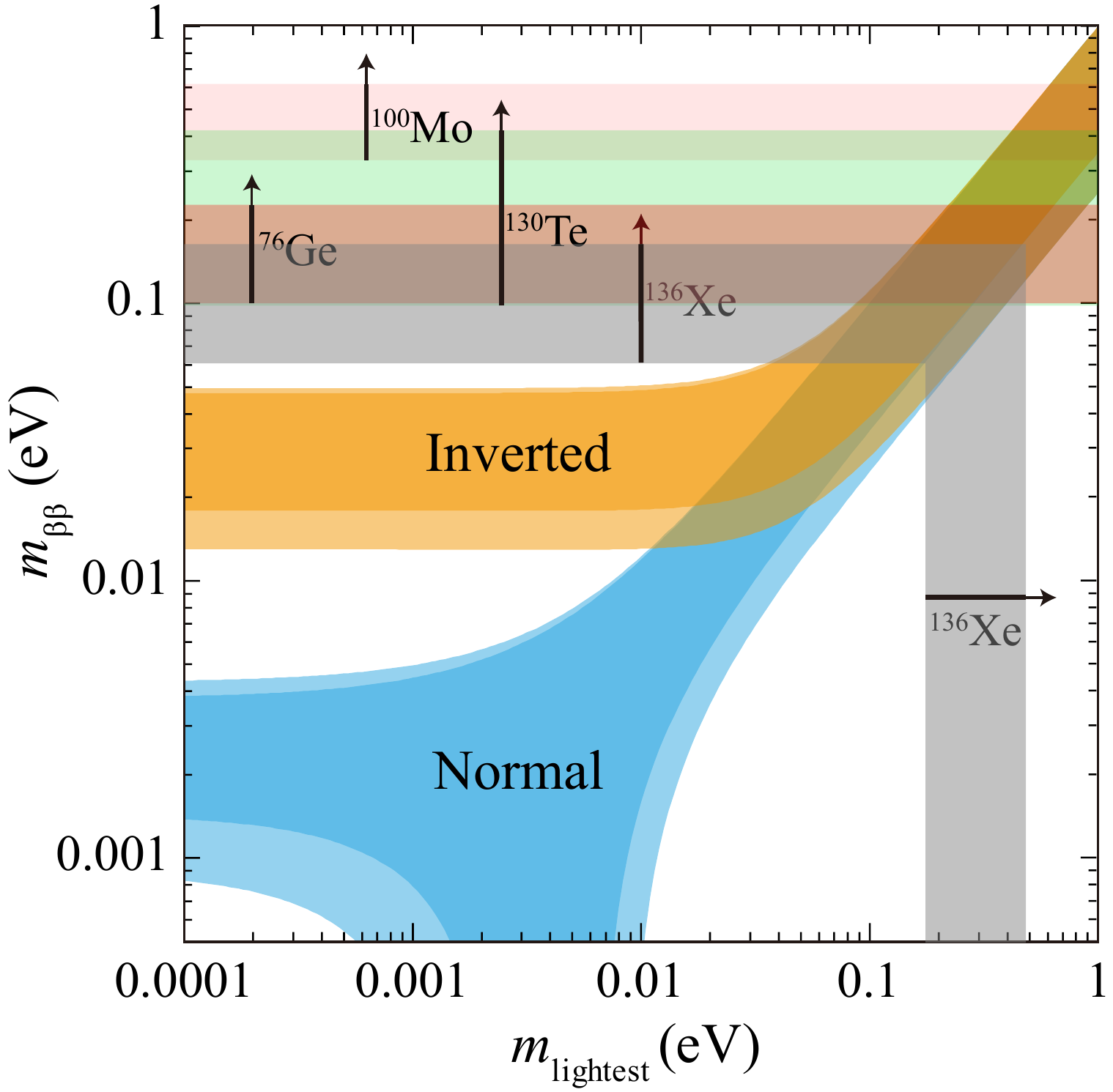}\hspace{2pc}%
\begin{minipage}[b]{17pc}\caption{\label{label} The allowed regions of the effective Majorana mass relative to the lightest neutrino mass in the normal and inverted orderings. 
The horizontal bands indicate the most stringent \mbb{} limits from \Ge{76}, \Mo{100}, \Te{130}, and \Xe{136} experiments from \cite{gerda2019taup}, \cite{nemo2015}, \cite{cuore2019taup}, and \cite{gando2016prl}, respectively. 
For the \Xe{136} result, the \mbb{} limit band confines a vertical region for the limit on $m_\mathrm{lightest}$.  The regions allowed for both orderings are adapted from~\cite{gando2016prl}.}
\label{m_bb_m_l}
\end{minipage}
\end{figure}

Because the $\nu$ oscillation measurements have yielded only the squared differences of the mass eigenvalues, 
which neutrino state is the lightest is not known.
Two possible cases of mass ordering exist. 
By setting $m_1$ or $m_3$ as the lightest of the mass eigenstates, the mass ordering becomes  either $m_1 < m_2 < m_3$ (normal ordering, NO) or $m_3 < m_1 < m_2$ (inverted ordering, IO), respectively.
The relation between \mbb{} and the mass of the lightest neutrino $m_\mathrm{lightest}$ can be established by using the parameters obtained in up-to-date studies of neutrino phenomenology. Fig.~\ref{m_bb_m_l} shows the allowed regions of the mass parameters in both ordering cases. The dark areas colored in light blue and in orange indicate allowed regions originating from the unknown CP phases for NO and IO, respectively. These allowed regions are widened when the experimental uncertainties on the parameters obtained from neutrino oscillation measurements are considered as shown by the light-colored regions. 
This plot indicates the general sensitivity of a \zerodbd{} experiment in the light $\nu$ exchange model. The half-life of \zerodbd{} is inversely proportional to $ m_{\beta\beta}^2$. Thus the smaller \mbb{} is the harder it will  be to discover \zerodbd. 
Notably, the allowed regions for \mbb{} and $m_\mathrm{lightest}$  are broadened not by the NME calculation uncertainties but by the unknowns and constraints related to the $\nu$ parameters.

In comparison with other experimental approaches, \zerodbd{} experiments provide the highest detection sensitivity for neutrino mass measurement. \zerodbd{} searches set the absolute mass scale of neutrinos in spite of model dependent decay mechanisms. 
On the other hand, a model independent approach to detect the neutrino mass is the precise measurement of the end point of a beta decay spectrum in KATRIN and ECHo projects~\cite{wolf2010nima,gastaldo2017epjc}. In such direct detection of neutrino mass, the mass observable is expressed as $m_{\beta}^2 = \sum |U_{ei}|^2 m_i^2$. The most up-to-date limit on $m_\beta$ is 1.1 eV from the KATRIN project~\cite{drexlin2019taup}.. 
Furthermore, observational cosmology constrains a strong but model-dependent limit on the sum of the neutrino masses, $m_\mathrm{tot} = \sum m_i$. 

Considering all the measured information on neutrino properties including \mbb, \ml, $m_\mathrm{tot}$, and neutrino oscillations, a global Bayesian analysis was performed to estimate the discovery probability distribution for \zerodbd~\cite{agostini2017prd}. This analysis yielded a probable region for the discovery rather than the allowed region indicated in Fig.~\ref{m_bb_m_l}. 
According to the prediction, if an experiment is carried out with a \mbb{} sensitivity of 20~meV, then the discovery potential for \zerodbd{} observation is expected to be 95\% in the case of IO. Even in the case of NO, the discovery potential becomes as high as 50\% with 20 meV detector sensitivity.  In NO, achieving 90\% chance of \zerodbd{} discovery will require 5 meV sensitivity. 

\section{Neutrinoless double beta decay experiments}

Although all the extant experiments have developed their own technologies for detecting the zero-neutrino mode of the decay, common strategies exist for increasing detection sensitivity. 
Greater exposure (i.e., the product of the amount of the $\beta\beta$ isotope and the measurement time) increases the chance of detecting rare events. Because the energy resolution defines the energy region of interest (ROI), higher resolution also improves the detection sensitivity in the presence of backgrounds. Finally, it is crucial to minimize the background events present in the ROI for a  \zerodbd{} search. 
These strategies have long motivated various technological developments in terms of detection methods, 
active and passive shields, purification and radio assays, laboratory controls/maintenance, and online/offline analysis.  

.  
A number of experiments worldwide have investigated and are currently investigating \zerodbd{} decays in several different nuclei
This review introduces several \zerodbd{} experiments based on commonly used isotopes, categorized in detector technologies.  Brief descriptions of present and future experiments are provided, including  presentations from the 16th International Conference of Topic in Astroparticle and Underground Physics (TAUP), Toyama, Japan, 2019. The most recent results are listed in Table.~\ref{table_mbb}.
There exist many other \zerodbd{} projects not covered in the review.

\subsection{Loaded liquid scintillators}
Loaded liquid scintillators (LSs) are a  popular choice of target materials in neutrino research. A large amount of additive acting as the target for neutrino interactions can be loaded into an organic LS with extremely low level of radioimpurities. 
Large exposure, a low background rate, and easy scalability are the evident advantages of using loaded LSs for \zerodbd{} searches.  

The present best limit on the \zerodbd{} rate is set by KamLAND-Zen using a LS hosted in a thin transparent balloon loaded with 400 kg of \Xe{136} gas~\cite{gando2016prl}. 
The half-life limit sensitivity and corresponding \mbb{} limit are \Tzerov{} $>$ 1.07$\times$10$^{26}$~years and \mbb{} $<$ (0.061--0.165)~eV, respectively. 
The gray region in Fig.~\ref{m_bb_m_l} indicates the upper limit of \mbb. The width of the limit band is attributed to the NME uncertainties~\cite{engel2017rpp}. The upper limit on the lightest neutrino has also been estimated to be \ml{} $<$ (0.18--0.48)~eV from the relation between \mbb{} and \ml{} as indicated by the vertical gray region in Fig.~\ref{m_bb_m_l}. 
A similar analogy can be applied to the other observables of neutrino mass, $m_{\beta}$ and $m_\mathrm{tot}$,  not discussed here. 
The KamLAND-Zen measurement has been resumed with 800 kg of \Xe{136} in 2019, and this experiment is expected to continue improving its sensitivity for \zerodbd{} discovery. 

SNO  was a heavy water Cherenkov detector that discovered the oscillation of $\nu_e$ to neutrinos of other flavors. Now, it has become SNO+, which is being filled with liquid scintillator. One of the science goals of  SNO+ is the search for \zerodbd{} of \Te{130}. Because of the large active volume, at 0.5\% \Te{\mathrm{nat.}} loading the detector contains 1.33 tons of \Te{130} and 20\% of that amount is in the fiducial volume~\cite{shimizu2019fp}. \Te{130} was chosen as the $\beta\beta$ isotope because of its high natural abundance of 34.1\% (see Table~1). %It will be the first 1 ton detector for \zerodbd{} observation. 
SNO+ expects to reach  a 2.1$\times$10$^{26}$~years sensitivity with a 5-year run~\cite{snoplut2019cossurf}.

\subsection{Ge semiconductors}
High-purity Ge (HPGe) detectors are a common class of radiation detectors for a variety of applications. Taking advantage of the mature technologies developed for commercial and research detectors, HPGe detectors consisting of isotopically enriched \Ge{76} can be prepared for a \zerodbd{} experiment. HPGe detectors provide the highest energy resolution (e.g., 2.5 keV FWHM at $Q_{\beta\beta}$ of \Ge{76}) among the technologies used in \zerodbd{} experiments~\cite{md2018prl}. %~\cite{avignone2019}. 
Moreover, the PSD features used to differentiate between bulk and surface events and between single-site and multi-site events are another powerful tool of Ge detectors for the \zerodbd{} search. 

The search for \zerodbd{} with Ge detectors started in the 1960s, and a number of experiments have contributed to continuing progress in sensitivity~\cite{avignone2019}. 
GERDA-II improved its background using an active veto consisting of a liquid Ar bath hosting Ge crystals with low-mass holder~\cite{gerda2018epjc}. This experiment found a lower limit of \Tzerov{} at 1.1$\times$10$^{26}$ years with \mbb{} $<$ (0.10--0.23)~eV~\cite{gerda2019taup}. 
Majorana Demonstrator employed the conventional configuration of Ge detectors in a vacuum cryostat and high-Z shielding. The copper used for the detector holder was electroformed and machined in the underground facility of their experimental site. A lower limit of \Tzerov{} at 2.7$\times$10$^{25}$ years was found in the Majorana Demonstrator~\cite{md2019epjc}.  

Recently, researchers involved in both GERDA and Majorana  initiated  the LEGEND collaboration. LEGEND will combine the strengths of the two groups, the low mass holder and active Ar veto of GERDA and radiopure materials and components of Majorana. 
 LEGEND-200 is to be prepared with 200 kg of \Ge{76} for a half-life limit  sensitivity of 10$^{27}$ years. 
 Eventually, a ton-scale detector will be built aiming at reaching a sensitivity of 10$^{28}$ years~\cite{legend2019taup}.

\subsection{Time projection chambers (liquid \& gas)}

The time projection chamber (TPC) is another technology of choice in \zerodbd{} experiments. 
TPCs, which can be built in liquid and gas phases, have several attractive features that are advantageous detecting $\beta\beta$ events at the Q value. 
Their scalability and self-shielding effect (e.g., in a liquid Xe TPC) are relevant.  
TPCs can also provide two detection channels  for ionization and scintillation signals allowing particle identification (PID). This two-channel detection provides competitive resolutions for energy, position and topology determinations. 

EXO-200, a single-phase liquid Xe TPC was built with an active Xe mass of 110 kg with 80.6\% \Xe{136} enrichment~\cite{exo2012ji}. The ionization and scintillation channels consist of wire grids and avalanche photodiodes, respectively. A total exposure of 234.1 kg$\cdot$year was completed resulting in limit sensitivities of \Tzerov{} $>$  3.5$\times$10$^{25}$ year and \mbb{} $<$ (0.093--0.286)~eV~\cite{anton2019}. The collaboration now plans to build nEXO, a 5-ton scale liquid Xe TPC reaching limit sensitivities of 
 9.2$\times$10$^{27}$ years and (0.009--0.018)~eV with a 10-year run~\cite{nexo2018prc}.  

Gas TPCs represent another promising technology for the \zerodbd{} search. 
A TPC with pressurized Xe gas has shown an  extrapolated energy resolution of 0.6\% FWHM at $Q_{\beta\beta}$ in the NEXT project~\cite{next2019taup}.  
NEXT-100, deployed with 100 kg of \Xe{136} at 15 bar, will reach  projected sensitivities of  9$\times$10$^{25}$ years  and (0.07--0.13)~eV~\cite{next2019taup}.

PandaX-III is another high pressure gas TPC being built using about 100 kg \Xe{136} for \zerodbd{} search~\cite{han2019}. It employs microbulk micromegas modules for the ionization channel. PandaX-III will reach a \Tzerov{} sensitivity of 9$\times$10$^{26}$ years in one module experiment of a 3-year run with possibility for multi-module detection.

   %and  & $<$0.376--0.770 & CUPID-0 & \cite{azzolini02018},

\subsection{Low-temperature thermal calorimeters}

Low-temperature (LT) thermal calorimeters operating at mK temperatures represent one of the promissing techniques used in rare-event experiments including \zerodbd{} searches. Such an LT detector uses a sensitive thermometer to measure the temperature increase of a crystal absorber caused by an input of energy to the system. Semiconductor thermistors (e.g., NTD Ge sensors), metallic magnetic calorimeters (MMCs), and transition edge sensors (TESs) are those thermometers used in thermal equilibrium detection providing a high energy resolution~\cite{enss2005cryogenic}. Various $\beta\beta$ decay isotopes can be chosen   as constitutive elements of absorber crystals for \zerodbd{} search. When a scintillating crystal is used as the absorber target, the light channel can be exploited in addition to the heat channel detection, enabling efficient PID. 

%As the present player in a series of 30 year advance using LT detectors for \zerodbd{} search, 
In a series of advances for searching \zerodbd{} using LT detectors for the last 30 years, 
it is CUORE that is searching  for \zerodbd{} of \Te{130} using 741~kg of TeO$_2$ crystals with NTD Ge thermistors \cite{alfonso2015prl}. The detector has an energy resolution of 5 keV FWHM at the \Te{130} $Q_{\beta\beta}$. Its \Tzerov{} sensitivity has reached  2.3$\times$10$^{25}$ years corresponding to \mbb{} $<$ (0.09--0.42)~eV \cite{cuore2019taup}. 

Using Zn$^{82}$Se scintillating crystals containing enriched \Se{82}, CUPID-0 carried out a \Se{82} \zerodbd{} search using the simultaneous detection method for heat and light signals  resulting in a limit sensitivity of  2.4$\times$10$^{25}$ years \cite{azzolini02018}. Similarly, CUPID-Mo exploits \LMO{} crystals in the heat and light detection resulting in a sensitivity of   3$\times$10$^{24}$ years \cite{cupidmo2019taup}. Recently, CUPID collaboration has formed a plan to search for \Mo{100} \zerodbd{} using \enLMO{} as the target crystal aiming at limit sensitivities of  \Tzerov{} $>$  1$\times$10$^{27}$ years and \mbb{} $<$ $\sim$0.02~eV \cite{cupid2019taup}.

AMoRE is another LT detector searching for \zerodbd{} of \Mo{100}. MMCs are employed as the sensor technology for the heat and light detection with high energy and timing resolutions~\cite{gbkim2017app}. A pilot stage of AMoRE with 1.9 kg of \enCMO{} composed of doubly enriched isotopes was completed resulting in a \Tzerov{} sensitivity of 9.5$\times$10$^{22} $\cite{amore2019epjc}. Presently, AMoRE-I is being prepared with 6 kg of \enCMO{} and \enLMO. Moreover, AMoRE-II has been fully funded, with plan of  building a 200 kg LT detector aiming at limit sensitivities of  \Tzerov{} $>$  5$\times$10$^{26}$ years and \mbb{} $<$ (0.017--0.029)~eV \cite{amore2019taup}.

\subsection{Other calorimeters}

NEMO-3 uses electron tracking calorimeters using thin source foils for many $\beta\beta$ isotopes including 1 kg of \Se{82}  and 7 kg of  \Mo{100}~\cite{nemo2011, nemo2015}. The most stringent  \Tzerov{} limits for \Mo{100} have been set by NEMO-3 (see Table~1). SuperNEMO is focused on searching for \zerodbd{} of \Se{82} using the tracking calorimeter method to reach a half-life sensitivity of 10$^{26}$ years for the present goal.  

CANDELS uses 305 kg of CaF$_2$ scintillating crystals placed in liquid scintillator to search for \zerodbd{} of \Ca{48}. Recently, CANDELS improved the sensitivity to 6.2$\times$10$^{22}$ years for  \zerodbd{} of  \Ca{48} \cite{candels2019taup}. 
Moreover, the group has started a low temperature experiment with CaF$_2$. A few successful measurements were carried out with CaF$_2$ crystals with MMC readouts for heat and light detection resulting in a clear PID capability \cite{candels2019taup}.

\begin{table}[t]
\caption{\label{ex} Measured and proposed limits on  $T_{1/2}^{0\nu}$ and \mbb{} from \zerodbd{} search experiments using commonly used isotopes, sorted by mass number and chronological order of publication.  }
\begin{center}
\begin{tabular}{c|l|llll}
\br
Isotope, Q$^a$, NA$^b$& $T_{1/2}^{2\nu}$$^c$ &$T_{1/2}^{0\nu}$& \mbb &Experiment &Ref. Year \\
~~~~~~  MeV,~  \%      &  10$^{19}$ y      & 10$^{24}$ y      & eV    &      &  \\
\mr
$^{48}$Ca, 4.268, 0.187&4.4$^{+0.6}_{-0.5}$ & $>$0.058& $<$3.5--22 &ELEGANT-IV & \cite{umehara2008}, 2008\\
											     &  & $>$0.062$^d$&   &CANDELS & \cite{candels2019taup}, 2019 \\
\mr
$^{76}$Ge, 2.039, 7.8  &165$^{+14}_{-12}$ & $>$90$^d$ & $<$0.10--0.23$^d$ &GERDA-II & \cite{gerda2019taup}, 2019 \\
							                       &  & $>27$ & $<$0.200--0.433 &Majorana D. &  \cite{md2019epjc}, 2019 \\
							                       &  & $>$1000$^e$ & $<$0.33--0.76$^f$ &LEGEND-200 &  \cite{legend2019taup}, 2021$^f$ \\	
							                       &  & $>$10$^4$$^e$ & $<$0.017$^e$ &LEGEND-1000 &  \cite{legend2019taup}, 2025/6$^f$ \\
\mr
$^{82}$Se, 2.998, 8.8  &9.2$\pm$0.7 & $>$0.36 & $<$0.89--2.43 &NEMO-3 & \cite{nemo2011}, 2011\\
										 &    &$>$100 & $<$0.05--0.1 & SuperNEMO & \cite{supernemo2017}, 2019$^f$\\
							            &    &$>$2.4 & $<$0.376--0.770 & CUPID-0 & \cite{azzolini02018}, 2018\\
\mr
%$^{96}$Zr, 3.356, 2.8  &2.3$\pm$0.2& -- & -- &NEMO-3 & --\\
%											          & -- & -- &    -& xxxx &\\
%\mr
$^{100}$Mo, 3.034, 9.7  &0.71$\pm$0.04 & $>$1.1 & $<$0.33-0.62  &NEMO-3 & \cite{nemo2015}, 2015\\
											          & & $>$0.095 & $<$1.2--2.1 & AMoRE-Pilot & \cite{amore2019epjc}, 2019\\
											          & & $>$10$^e$ & $<$0.12--0.2$^e$ & AMoRE-I & \cite{amore2019taup}, 2019$^f$\\											          
											          & & $>$500$^e$ & $<$0.017--0.029$^e$ & AMoRE-II & \cite{amore2019taup}, 2021$^f$\\											          
											          
											          &  & $>$0.3$^d$ &$<$0.715--1.19$^d$& CUPID-Mo & \cite{cupidmo2019taup}, 2019 \\
											          &  & $>$1000$^d$ &$<$20$^d$& CUPID & \cite{cupid2019taup}, 2025$^f$ \\
%\mr
%$^{116}$Cd, 2.802, 7.5  &2.87$\pm$0.13 & -- & -- &NEMO-3 & --\\
\mr
$^{130}$Te, 2.528, 34.1  &69$\pm$13 & $>$23$^d$ & $<$0.09--0.42$^d$ &CUORE & \cite{cuore2019taup}, 2019\\
											          &  & $>$210$^e$ &$<$0.03$^g$& SNO+ & \cite{snoplut2019cossurf}, 2020$^f$ \\	
\mr
$^{136}$Xe, 2.458, 8.9  &219$\pm$6 & $>$107 & $<$0.061--0.165 &KamL.-Zen400 & \cite{gando2016prl}, 2017\\
											      &   &$>$500$^e$  & $<$0.028--0.076$^e$ & KamL.-Zen800 &\cite{kamdland2019taup}, 2019\\
											      &   &$>$35  & $<$0.093--0.286 & EXO-200 &\cite{anton2019}, 2019\\
											      &   &$>$920$^{e}$  & $<$0.009--0.018$^{e}$ & nEXO &\cite{saldanha2019taup}, 2027$^f$\\
											      &   &$>$0.21  & $<$1.4--3.7 & PandaX-II &\cite{ni2019cpc}, 2019\\
											      &   &$>$90$^e$  & $<$0.06--0.18$^e$ & PandaX-III &\cite{han2019}, 2020$^f$\\
	     									      &   &$>$90$^e$  & $<$0.07--0.13$^e$ & NEXT-100 &\cite{next2019taup}, 2020$^f$\\
%\mr
%$^{150}$Nd, 3.371, 5.6  &0.82$\pm$0.09 & -- & -- &NEMO-3 & --\\
%											         & -- & -- &    -& AMoRE-Pilot & --\\
\br
\end{tabular}
\end{center}
$^a$~Q values from \cite{wang2017ame2016}.
$^b$~Natural abundances \cite{meija2016}.
$^c$~Recommended values from \cite{barabash2015}. 
$^d$~Limits adapted from the corresponding TAUP presentations, as listed in the Ref. column. 
%$^e$~Proposed limit setting sensitivity from presentations in TAUP2019. 
$^e$~Proposed limit-setting sensitivities for ongoing or future experiments. 
$^f$~Proposed schedules for data taking, subject to change. 
$^g$~Value from \cite{shimizu2019fp}.
\label{table_mbb}
\end{table}

\section{Conclusion}

%The technologies of the \zerodbd{} experiments are well-developed. 
%Many experiments with well-established technologies carry out careful but promising approaches to probe the double beta decay in zero-neutrino mode. 
Many experiments are using well-established technologies to carry out careful but promising approaches to probe the possibility of double beta decay in the zero-neutrino mode.
It will be important to reach a \mbb{} sensitivity of  20~meV to achieve 90\% discovery potential in IO scenario. 
Doing so will prove or disprove that $m_3$ is the lightest one of the Majorana neutrinos. 
Moreover, even if the NO case holds true, as suggested by recent results from a few experiments that favor NO at 3$\sigma$, this sensitivity will provide a 50\% probability of discovering \zerodbd. 
Several experiments have been either funded or proposed that are expected to cover the sensitivity region near 20 meV in the next decade. 

%Doing so will allow the claim that $m_3$ is the lightest of the Majorana neutrinos to be either proven or disproven.
%Moreover, even if it is in the case of  NO as recent results from a few experiments favor NO in about 3$\sigma$, the sensitivity provides 50\% discovery probability  to discover \zerodbd.  
%Several experiments have been either funded or proposed  covering the sensitivity region near 20 meV in the following decade.  

 \zerodbd{} discovery would be a clear demonstration of Majorana neutrinos and the existence of LNV mechanisms. To clarify the physics mechanism responsible for \zerodbd, many experiments with various isotopes and different detection methods will be required. Research on \gA{} and NMEs will gain advantages from studies with many isotopes.  
 In this sense, \zerodbd{} experiments are not just a neutrino mass experiment but also 
also provide an avenue of studying neutrino mass to gain a fundamental understanding of 
particle physics.

\section*{Acknowledgments}
This work was presented in the plenary session of TAUP 2019. It was supported by Grant No. IBS-R016-G1.
\section*{References}
%\bibliographystyle{iopart-num-simple}
%\bibliography{dbd_ref_taup}

\providecommand{\newblock}{}

\end{document}